\documentclass[reprint,prl,aps, floatfix]{revtex4-1}
\usepackage{graphicx}
\usepackage{color}
\usepackage{dcolumn}
\usepackage{bm}
\usepackage{amssymb}
\usepackage{braket}
\usepackage{color,amsmath}
\usepackage{comment}
\usepackage{gensymb}
\usepackage{soul,xcolor}
\setstcolor{red}
\usepackage{hhline}
\usepackage{tabularx}
\usepackage{xspace}
\usepackage{layouts}
\usepackage{lineno}
\usepackage{booktabs}
\usepackage{longtable}
\usepackage{tcolorbox}
\tcbuselibrary{skins,breakable}
\usepackage[hyperfootnotes=false,colorlinks=true,allcolors=blue]{hyperref}
\usepackage{xr}
\newcolumntype{C}[1]{>{\centering\arraybackslash}m{#1}}
\newcolumntype{N}{@{}m{0pt}@{}}

\providecommand{\unit}[1]{}\renewcommand{\unit}[1]{\,\mathrm{#1}}
\providecommand{\equa}[1]{}\renewcommand{\equa}[1]{Eq.~\eqref{#1}}

\newtcolorbox{aiscientistbox}{
    breakable,
    colback=gray!10,
    colframe=gray!80,
    coltitle=white,
    colbacktitle=gray!70,
    fonttitle=\bfseries,
    title=AI agent output,
    boxrule=0pt,
    sharp corners,
    enhanced,
    attach boxed title to top left={yshift=-2mm, xshift=2mm},
    boxed title style={sharp corners, size=small, boxrule=0pt}
}

\newtcolorbox{promptbox}{
    breakable,
    colback=gray!10,
    colframe=gray!80,
    coltitle=white,
    colbacktitle=gray!70,
    fonttitle=\bfseries,
    title=Prompt,
    boxrule=0pt,
    sharp corners,
    enhanced,
    attach boxed title to top left={yshift=-2mm, xshift=2mm},
    boxed title style={sharp corners, size=small, boxrule=0pt}
}
\begin{document}

\title{AIMS: an AI experimentalist turns uncertainty into quantum matter discovery}

\author{Siyuan Qiu$^{1,*}$}
\author{Philip D. Suh$^{1,*}$}
\author{Nhat Huy Tran$^{5}$}
\author{Xirui Wang$^{7,9}$}
\author{Heonjoon Park$^{8}$}
\author{Kutay Akin$^{1}$}
\author{Kevin K. S. Multani$^{1,7,9}$}
\author{Seungwon Jung$^{2,3}$}
\author{Wenkai Cai$^{2}$}
\author{Xinyu Liu$^{2,10}$}
\author{León Garcia$^{1}$}
\author{Ziyan Zhu$^{6}$}
\author{Chunjing Jia$^{5}$}
\author{Zhantao Chen$^{4,11}$}
\author{Zhixun Shen$^{1,7,8,9,\dagger}$}
\author{Zhurun Ji$^{2,9,\dagger}$}

\affiliation{$^{1}$ Department of Physics, Stanford University, Stanford, CA 94305, USA}
\affiliation{$^{2}$ Department of Physics, Massachusetts Institute of Technology, Cambridge, MA 02139, USA}
\affiliation{$^{3}$ Department of Materials Science and Engineering, Massachusetts Institute of Technology, Cambridge, MA 02139, USA}
\affiliation{$^{4}$ Walker Department of Mechanical Engineering, The University of Texas at Austin, Austin, TX 78712, USA}
\affiliation{$^{5}$ Department of Physics, University of Florida, Gainesville, FL 32611, USA}
\affiliation{$^{6}$ Department of Physics, Boston College, Chestnut Hill, MA 02467, USA}
\affiliation{$^{7}$ Department of Applied Physics, Stanford University, Stanford, CA 94305, USA}
\affiliation{$^{8}$ Geballe Laboratory for Advanced Materials, Stanford University, Stanford, CA 94305, USA}
\affiliation{$^{9}$ Stanford Institute for Materials and Energy Sciences, SLAC National Accelerator Laboratory, Menlo Park, CA 94025, USA}
\affiliation{$^{10}$ Department of Physics, New York University, New York, NY 10003, USA}
\affiliation{$^{11}$ Texas Materials Institute, The University of Texas at Austin, Austin, TX 78712, USA}
\affiliation{$^{*}$ These authors contributed equally to this work.}
\affiliation{$^{\dagger}$ Correspondence: zxshen@stanford.edu, zhurun@mit.edu}

\maketitle

\textbf{Most AI agents act only after scientists have defined the task. Discovery is harder under practical uncertainties: the probe may not be where it is expected, the signal may occupy only a small region of a disordered sample, and the evidence may not distinguish among competing explanations. Here we show that an AI agent can decide what evidence an uncertain experiment needs next, and act on it. Beyond automation, AIMS, an uncertainty-aware experimentalist for cryogenic microwave impedance microscopy, quantifies uncertainty where it originates, in perception, sampling, and interpretation, and converts each into its own corrective action rather than a single confidence score. Given only an open objective, AIMS relocated a probe lost during cooldown while flagging its own unreliable estimates, mapped twist angle disorder to locate the strongest correlated states in twisted bilayer MoSe$_2$, and uncovered a paradox: the half-filled stripe that classical theory predicts should melt first survived longest. Distinguishing an incomplete model from a wrong mechanism, AIMS commissioned a beyond-mean-field calculation and an independent structural measurement as the decisive tests, revising its interpretation as each arrived: quantum motion reverses the classical hierarchy, stabilizing the half-filled stripe while destabilizing its neighbors. These uncertainty-to-action loops are generic to scanning probe experiments, and AIMS turns uncertainty from an obstacle into a driver of discovery.}

\section{Introduction}
AI-powered systems are beginning to automate scientific research across domains~\cite{wang_scientific_2023} --- from chemical synthesis~\cite{boiko_autonomous_2023, bran_augmenting_2024, dai_autonomous_2024, ramos_review_2025} and materials optimization~\cite{ni_matpilot_2024} to self-driving laboratories~\cite{macleod_self-driving_2020, szymanski_autonomous_2023, tom_self-driving_2024, zhang_multimodal_2025}. Built on reasoning-capable large language models that plan, use tools and iterate~\cite{wei_chain_2022, yao_react_2023}, such agents increasingly act on scientific instruments as well --- executing gene-editing experiments~\cite{qu_crisprgpt_2026}, operating advanced instruments that adapt on the job~\cite{vriza_operating_2026}, and running multistage experiments at large-scale facilities~\cite{hellert_agentic_2026, chen_agentic_2025}. Yet most operate where the inputs are digital or highly structured, the action space is predefined, and success is judged against a fixed objective. At the frontier of this progression, an LLM-driven end to end closed loop AI experimentalist has been demonstrated to autonomously fabricate van der Waals materials and assemble into 2D devices~\cite{shi2026qumus}. This invites an important question in any real experiment where uncertainty is inevitable: whether an AI agent can work inside a physical experiment where observations are noisy, useful measurements are sparse, and the scientific interpretation itself remains undetermined.

Quantum materials experiments expose this challenge sharply. A cryogenic scanning probe experiment is not only slow or technically demanding; it is uncertain at every level. The probe may not be where it is expected to be after cooldown, the relevant electronic phase may occupy only a small region of a spatially inhomogeneous device \cite{Uri2020TwistAngleDisorder, Kazmierczak2021StrainFields, Li2024FlatBandRelaxations, Hoke2026SupermoireRelaxation, Sung2020BrokenMirrorSymmetry, Shabani2021DeepMoirePotentials}, and even with high-quality data, the microscopic mechanism may remain under-determined, because classical, quantum and disorder-related explanations produce overlapping signatures~\cite{Shayegan2022WignerCrystalsFlatBand, Matty2022MeltingGWC, Zhou2024QuantumMeltingGWC, Nuckolls2024MicroscopicPerspectiveMoire, Kumar2026MeltingTemperatureGWC}. The relevant benchmark here is therefore not autonomy alone, but uncertainty-aware experimental agency: whether the agent recognizes unreliable perceptions, changes its measurement plan when the sample is disordered, and preserves competing mechanisms when the data cannot yet select a single explanation.

Microwave impedance microscopy (MIM) \cite{lai2010mesoscopic, ma_mobile_2015, cui_unconventional_2016, allen_visualization_2019, barber_microwave_2021} provides a stringent test of this standard. Cryogenic MIM has revealed local electronic structure in quantum Hall, quantum spin Hall, quantum anomalous Hall and fractional Chern systems \cite{cui_unconventional_2016, shi_imaging_2019, allen_visualization_2019, ji_local_2024}, but its power comes with the same obstacles that frustrate autonomy in many physical laboratories. Thermal contraction can displace the tip by hundreds of microns, local disorder and twist-angle variation obscure the best measurement region, and the interpretation of correlated insulating states demands comparison across energy scales, models and structural perturbations. A useful agent for this setting must therefore do more than automate scanning: it must convert uncertainty into the next physical or analytical action.

Here we introduce AIMS, a closed-loop large-language-model (LLM) agent that operates a cryogenic MIM platform end to end. Rather than pursuing autonomy as an end in itself, AIMS is designed to make uncertainty actionable: three nested loops carry the experiment from measurement preparation, through data accumulation, to physical conclusion, each recognizing its dominant uncertainty and converting it into the next measurement or calculation. The result is an uncertainty-aware AI experimentalist for quantum matter — autonomous where the task is well posed, collaborative where scientific judgment is high-stakes, and explicit about the evidence needed to reduce remaining ambiguity.

\section{AIMS: an uncertainty-aware AI experimentalist}

As detailed in \textbf{Figure~1}, AIMS confronts the three uncertainties that pervade a MIM experiment — \textit{Where am I? Where should I measure? What physics explains the data?} — each converted into the next concrete action by a dedicated loop.

The navigation loop confronts uncertainty in perception of tip position. Prompted to locate the sample, AIMS scans through MCP server \cite{modelcontextprotocol2025specification} encoded instrument control and predicts the tip position by automated localization, moving and re-scanning until the tip reaches the target, keeping human involvement minimal with explicit self-reasoning and reliable failure handling tools.  

The measurement loop confronts uncertainty in the inhomogeneous distribution of correlated states. A large-area scan swept across gating voltage leads to isolation of strongly correlated regions; grid spectroscopy then maps local properties such as twist angle and disorder, steering AIMS toward low-disorder, high-contrast sites with scoring of each site by the quality of its correlated features. Uncertainty about sample quality is thereby converted into a ranked landscape that pins down the optimal measurement position, with researchers free to cross-check the agent's choice.

The discovery loop confronts uncertainty in interpretation, where classical and quantum mechanisms produce overlapping signatures that no single measurement can disentangle. Rather than forcing a binary verdict, AIMS treats interpretation as an evidence-updating process. Guided by prompts encoding the analysis workflow, it reviews its database of experimental data, literature and tools, and ranks both researcher-seeded and self-generated hypotheses by data-derived evidence rather than committing prematurely to one model. When the available evidence is insufficient, AIMS identifies the specific gap, proposes the follow-up calculation or measurement most likely to resolve it, and incorporates the resulting evidence into an updated mechanism ranking. Researchers provide the guiding question, initial hypotheses, and experimental safeguards, while AIMS expands the hypothesis space, ranks competing mechanisms, specifies follow-up tests, and updates the physical attribution.

Across all three loops, performance was evaluated against independent controls rather than successful task completion alone. Navigation was repeated from different initial positions and on the actual cryogenic device; measurement selection was cross-checked using an independent image metric and a denser human-defined grid; and mechanism attribution compared matched classical and quantum forward models and was challenged by beyond-mean-field calculations and registered structural measurements. The resulting analyses were further evaluated blindly against human reports. The supplied inputs, autonomous decisions, uncertainty triggers, human interventions, and success criteria are summarized in \textbf{SI Table S2}.

\section{Locating sample under uncertain perception of position}

To benchmark AIMS-driven closed-loop navigation, we first tested it on a patterned SiO$_2$/Si marker chip with densely distributed triangular marks encoding directional information toward the sample. Rather than place an actual sample, we defined successful navigation as bringing the tip within 150 $\mu$m of the chip center, where sample would usually reside, and initiated each test by displacing the tip $\sim$500 $\mu$m in a random direction, comparable to the offset typical after a cooldown. Following the closed-loop workflow, AIMS drove the tip to within 150 $\mu$m of the center in six scans (\textbf{Figure~2(d)}). The stark conductivity contrast between the Au markers and the insulating substrate renders the markers clearly visible in the MIM-Im channel, enabling position prediction by particle filter and template matching \cite{arulampalam2002tutorial, lewis1995fastnormalized, dalal2005histograms} with the optical image, which outperforms convolutional neural network (CNN) method \cite{krizhevsky2012imagenet} in accuracy. 

Even with careful marker design, disturbances under practical measurement conditions can still corrupt individual predictions. For example, the predicted position for Scans 3 is visibly inaccurate as a result of poor image quality (\textbf{Figure~2(e)}). Crucially, AIMS recognizes this unreliable prediction through its large uncertainty and flags it in its output, correctly explaining that the tip retracted on encountering surface debris during both scans (\textbf{SI Figure S12}). AIMS thus not only initializes the navigation loop but also diagnoses the origin of experimental failures as feedback.

Such defects are endemic to two dimensional devices and persistently degrade SPM image quality. AIMS was therefore equipped with recovery strategies — re-scanning at increased tip lift, or Gaussian process (GP) reconstruction\cite{kelley_fast_2020, ziatdinov_imaging_2020, fu_accelerated_2024} of a higher-resolution image with enhanced contrast (\textbf{Figure~2(f)}) — that it can invoke autonomously when uncertainty is elevated. The application of GP regression on Scan 3, for instance, brought the prediction back to the correct quadrant and sharply suppressed the uncertainty. Compared to human-driven navigation with scripted code for instrumentation control, we found the AIMS-driven workflow reduced total navigation time from $\sim$10 hours to $\sim$4 hours, owing principally to fewer required scans under more confident position estimation and a more progressive movement policy.

We further validated the workflow from different initial positions (\textbf{SI Figure~S1}) and after cooldown to cryogenic temperature on a real device (\textbf{SI Figure~S2}), where AIMS autonomously invoked GP regression to avert false predictions caused by low image quality. AIMS thus functions not as a passive measurement tool but as an active closed-loop problem solver that reasons over uncertainty and adapts its strategy to practical experimental challenges.

\section{Measuring correlated states under sample inhomogeneity}

After the sample location is identified, the measurement loop confronts a different uncertainty: which experimental conditions, across a large parameter space — position, gating voltage, magnetic field, temperature and other degrees of freedom — will expose the physics of interest. To explore this, we fabricated a twisted bilayer MoSe$_2$ device (Methods), targeting a 4$^\circ$ twist angle, previously reported to exhibit Mott insulating state at integer filling $\nu=1$ and generalized Wigner crystal (GWC) states at fractional fillings \cite{zong2025quantum}. However, because transfer uncertainties and substrate-induced strain can drive the experimental twist angle substantially from target and introduce spatial non-uniformity, the productive region must be surveyed rather than presumed.

By looping between data acquisition and analysis, AIMS progressively narrowed the optimized measurement region from micron scale to a few candidate points, as illustrated in \textbf{Figure~3(a)}. AIMS began with a large scale $3\,\mathrm{\mu m}\times 3\,\mathrm{\mu m}$ MIM scan while sweeping bottom gate voltage, focusing on the Mott state. The smaller region of stronger correlated response is then extracted by calculating the cumulative pixel-to-pixel difference between each pair of MIM-Im images acquired at adjacent gate voltages, where the appearance of correlated insulating state leads to suppression of local conductivity, as in \textbf{Figure~3(b-c)}. As an independent check by human decision, a simultaneous pixel-wise comparison using structural similarity index (SSIM) is performed, which shows a well-overlapped correlated region (\textbf{SI Figure~S5(a)}), supporting AIMS's choice. The SSIM method was subsequently added to AIMS as an additional MCP tool for correlated region identification.

Within the selected region, AIMS set up a grid spectroscopy measurement(\textbf{Figure~3(e)}) and revealed clear spatial variation with shifting correlated dip positions that signal substantial twist angle disorder. Assigning the most prominent dip to the $\nu=1$ Mott state, AIMS produced an in-situ twist angle disorder map (\textbf{Figure~3(f)}), showing an average twist angle $\theta=3.13^\circ\pm0.22^\circ$. 

Beyond mapping where twist angle inhomogeneity lies, AIMS pins down the best measurement position by scoring the quality of the correlated features measured at each grid point — weighing the number of fractional states, the depth of each dip, the measurement noise, and the deviation from the predicted fractional fillings (\textbf{Figure~3(g)}), turning a subjective site choice into a reproducible ranking. On this basis, AIMS selected the point at row 3, column 9 (star in \textbf{Figure~3(f)}), which resolves a series of fractional GWC states at $\nu$= 1/3, 1/2, and 2/3, alongside the $\nu=1$ Mott insulator and $\nu=2$ band insulator. (\textbf{Figure~3(h)}). The independent human benchmark investigates a larger grid area (\textbf{SI Figure S5(b-f)}), partially overlapping AIMS's region and taken under more finely tuned tip conditions, but still converged on a best position within $\sim$200 nm of AIMS's choice, further corroborating its choice without human intervention. Overall, the measurement loop demonstrates an improvement in resolving uncertainties in measurement position of real sample automatically, combining the mapped twist angle disorder with a quantitative ranking of correlated feature quality, and feeds both back to researchers who remain free to verify the outcome.

\section{Resolving generalized Wigner crystal melting mechanisms under ambiguous interpretation}

The melting of generalized Wigner crystals \cite{Regan2020, Xu2020, Huang2021FractionalWS2WSe2, zong2025quantum, Yang2025CorrelatedInsulating} presents an intriguing physical problem. In twisted MoSe$_2$, the fractional states at $\nu=1/3, 1/2$, and $2/3$ arise from the same moiré system, yet their thermal stability differs sharply. Temperature-dependent MIM spectroscopy shows that the $\nu=1/3$ and $2/3$ states melt near 17 K, while the $\nu=1/2$ stripe remains robust throughout the accessible range, establishing a melting hierarchy $T_{\mathrm m}(\nu=1/2)>T_{\mathrm m} (\nu=1/3)\simeq T_{\mathrm m}(\nu=2/3)$. The extrapolated value $T_{\mathrm m}(\nu=1/2)=30.5\pm6.0$ K is treated as censored evidence, since the feature remains finite at the 20 K measurement ceiling, but the hierarchy itself is directly established (\textbf{Figure~4(b)}). The central question is therefore why the half-filled stripe  resists thermal melting far more than its neighbors.

Experimental interpretation commonly proceeds by settling on the single model that best matches the data, but this can leave several questions entangled: a model may capture the defining trend while remaining quantitatively incomplete; a residual may reflect missing physics rather than a competing mechanism; and sample heterogeneity may control one observable without causing the primary phenomenon. A binary classical-or-quantum verdict does not, by itself, distinguish these possibilities.

We thus built the discovery loop of AIMS to make this interpretive process itself closed-loop (\textbf{Figure~4(a)}). Starting from a scientific question and researcher-seeded hypotheses, AIMS generates additional explanations from model residuals, measurement limitations, and sample heterogeneity, then separately evaluates support for each physical mechanism, the possibility of an alternative mechanism or hidden variable, and the quantitative closure of the current model. Each unresolved uncertainty is converted into a specific action—calculation, characterization, calibration, or measurement—and the resulting evidence is incorporated into a new inference pass. Here, researchers supplied two initial hypotheses, classical thermal and quantum-fluctuation-influenced melting, which AIMS independently expanded with twist-domain inhomogeneity, the MIM dip depth as an imperfect proxy for thermodynamic melting, and beyond-mean-field physics as open mechanisms.

AIMS's first pass established that classical model alone cannot produce the observed hierarchy. Across the explored dielectric-parameter space, a hopping-free ($t=0$) Monte Carlo (MC) model consistently predicted the $\nu=1/2$ state to be the least robust, opposite to experiment (\textbf{Figure~4(c)}). On the contrary, a finite-hopping ($t\neq0$) Hartree–Fock (HF) model reproduced the $\nu=1/2$-dominant ordering under identical priors and likelihoods, with parameter attribution identifying Coulomb repulsion and hopping as leading controls of the calculated melting scales (\textbf{Figure~4(d)}). The data therefore favored a quantum-fluctuation origin.

Crucially, AIMS did not treat the first-pass model ranking as a final verdict. The favored mean-field explanation retained high model-closure uncertainty $U_\text{closure}$, particularly for the over-stablized $\nu$=2/3 melting scale. In addition, the measured twist-angle variation raised a structural uncertainty: local morphology might drive the melting hierarchy or might instead control only the spatial strength of the correlated response. These two uncertainties led to two actions: exact diagonalization (ED) and finite-temperature Lanczos method (FTLM) calculations to test the mechanism beyond mean field, and an AFM measurement registered to the MIM data to test the structural hypothesis. The first inference pass thus generated not only a leading explanation, but also the evidence required to challenge and refine it.

The ED/FTLM follow-up strengthened the quantum mechanism assignment and revealed the microscopic origin of the hierarchy. Increasing hopping keeps $\nu=1/2$ the most robust state while further suppressing the stability of $\nu=1/3$ and $2/3$ (\textbf{Figure~4(e)}), agreeing with experiment and thus reducing $U_\text{closure}$. Hopping lowers the melting scale of the $\nu=1/3$ and $2/3$ crystal, consistent with quantum fluctuations assisting thermal disordering. At $\nu=1/2$, however, the same kinetic term strongly raises the melting scale, from approximately 5 K near the small-hopping limit to approximately 30 K at the physical hopping. The exceptional thermal stability of the $\nu=1/2$ stripe is thus not simply inherited from stronger classical charge order but enabled by electron hopping. The results therefore show that quantum motion can act in opposite directions on the competing charge configurations.

The registered AFM measurement resolved the structural uncertainty. Registered topography and twist angle analysis showed two twist domains separated by a topographic fold (\textbf{SI Figure S8(a)}). The temperature-series point lies in a homogeneous high-twist domain, so twist disorder does not account for the measured melting hierarchy. Instead, the fold controls a different observable: the spatial strength of correlated features, which correlates directly with distance from the fold (\textbf{Figure~4(f)}). Thus, local morphology organizes where correlated states are strongest, but does not replace the intrinsic mechanism of the fractional melting hierarchy, allowing AIMS consequently reassigned the structural hypothesis as a secondary mechanism.

In the second inference pass, AIMS incorporated both follow-ups into the same mechanism-ranking framework, further favoring quantum-fluctuation-influenced melting as the primary mechanism for the fractional hierarchy with reduced $U_\text{closure}$ while assigning twist disorder and other alternatives a secondary role (\textbf{Figure~4(g)}). The resulting picture assigns distinct roles to the relevant processes: interactions provide the basis for charge crystallization; quantum kinetics reconstructs the filling-dependent thermal hierarchy; and morphology controls where GWC responses are strongest. This closes the discovery loop. AIMS transforms an initially binary question into a quantitative account by expanding the hypothesis space, separating distinct uncertainties, generating the evidence needed to resolve them, and updating the interpretation when that evidence arrives. The resulting account reveals quantum-enabled thermal stability at half filling while preserving the remaining under-stabilized $\nu=2/3$ state predicted by ED/FTLM and morphology-dependent spatial response as targets for further investigation. In a blind evaluation using the same dataset and tools, independent condensed-matter physicists scored the resulting AIMS reports comparably to the strongest human analyses (\textbf{Figure~4(h)}).

\section{Discussion and Outlook}
In conclusion, AIMS demonstrates an uncertainty-aware closed-loop agency in a cryogenic scanning probe quantum materials experiment, coupling perception recovery, adaptive measurement selection, and mechanism attribution in one complete experimental workflow. The navigation loop frees researchers from redundant but mandatory steps of the experiment. Moreover, the adoption of agent intelligence and reasoning capability allows it to recover from experimental failures and operate robustly under uncertainties. The measurement loop on the twisted MoSe$_2$ sample further manifests the use of AIMS to unveil hidden physical inhomogeneities, such as twist angle disorder, and optimizes the experimental parameters based on quantitative correlated feature evaluation to improve the efficiency and quality of experiments. Building on these data, the discovery loop explores classical and quantum hypotheses of Wigner-crystal melting, uncovering a quantum-fluctuation-renormalized picture of the anomalously robust $\nu=1/2$ crystal with solid evidence-backed inference process and proposed future experiments closing the three nested loops.

Several limitations temper these results and define the work that remains. The demonstration of AIMS centers on a single cryogenic instrument and material system. Although the loop architecture is modality-agnostic by design, where new capabilities enter simply as additional tools within the same stateful server, the generalization across the wider scanning probe family and samples exhibiting different correlated behaviors awaits experimental test. Furthermore, the parameter-space exploration focuses only on position, without full investigation on gate voltage, magnetic field, and temperature yet. Lastly, the current discovery loop still depends on curated tools, literature, and human-seeded questions, and the mechanism attribution remains probabilistic and evidence-ranked, not a final microscopic theory. We anticipate that coupling the discovery loop more tightly to versatile theory, allowing the agent to pose its own scientific questions and orchestrate measurements across successive cooldowns, would finally close the remaining gap toward genuinely self-driving condensed matter physics discovery beyond uncertainties. 

More broadly, AIMS points to a different criterion for evaluating scientific agents in quantum matter. Full autonomy is not, by itself, the relevant benchmark for a cryogenic experiment in which perception can fail, the sample is intrinsically inhomogeneous, and microscopic interpretation remains under-determined. The more meaningful standard is uncertainty-aware evidence generation: whether the agent recognizes when its own position estimate is unreliable, adapts its measurement plan when the sample is disordered, and preserves competing mechanisms when the data cannot yet select one. In this experiment, uncertain perception triggered recovery, sample inhomogeneity triggered adaptive measurement selection, and ambiguous interpretation triggered targeted theoretical and structural follow-ups — converting uncertainty at the instrument, sample, and mechanism levels into concrete scientific action. This standard also bridges experiment and theory: experimentalists confront drift, defects and sparse signal, while theorists confront competing models and uncertain energy scales, and AIMS connects both in one loop, turning experimental uncertainty into better measurements and theoretical ambiguity into explicit, testable hypotheses. Within this architecture, graded human involvement makes AIMS trustworthy — autonomous where the task is well posed, collaborative where scientific judgment is high-stakes, and explicit about the measurements needed to reduce remaining ambiguity. Such behavior distinguishes an AI experimentalist from a scripted automation pipeline and offers a practical framework for AI-assisted discovery in physical laboratories.

\section*{Methods}
\label{sec:methods}
\noindent\textbf{AIMS implementation}
The navigation and measurement loops of AIMS were built on Claude Opus 4.5, accessed through LibreChat using application programming interface (API) access, with instrument control and data analysis capabilities exposed as tools through a Model Context Protocol (MCP) server. Representative MCP tools are summarized in \textbf{SI Table 3}. The model was used with its default configuration, except a long timeout of 300000~s to prevent timeout-induced interruptions during scanning. Internet access was disabled to prevent the agent from taking unnecessary or uncontrolled coding actions that could interfere with the experiment, while preserving its ability to identify and handle experimental uncertainties.
The agent was prompted to continue whenever a loop was interrupted, for example running out of API token credits. 

Each run was initialized with a task prompt that specified the essential run-specific information, including the working directory, parameters for scanning, dimensions of templates, and other experimental settings, following similar operation logic of human-driven measurements. The intrinsic uncertainties of the two loops were autonomously identified by AIMS without being explicitly highlighted in the prompt, although possible mitigation strategies were included to guide AIMS in overcoming these difficulties. Two prompts for the navigation and measurement loops are shown in the Supplementary Information as examples. 

The discovery loop of AIMS was constructed on Claude Opus 4.8, accessed via the Claude Desktop interface with full Internet access, reflecting the higher level of freedom in programming beyond simply calling the provided MCP tools. All agent-directed computation was executed locally on a single CPU-only workstation (Intel Xeon-class processor, Sapphire Rapids, Family 6 Model 143; 12 physical cores / 24 logical threads; 68 GB RAM) running Windows 11 (AMD64) with Python 3.12.7; no GPU was used. The prompt is separated into two parts with different focuses. The physics prior prompt includes the documentation for each MCP tool, data and literature, together with details of the fabrication parameters of the device, helping AIMS to gain a more comprehensive understanding of the database. On the other hand, the Bayesian inference prompt instructs AIMS to address the mechanism attribution as a calibrated, updateable inference procedure rather than a qualitative judgment with scalable posterior probabilities and Bayes factors produced by code. Whenever a mechanism remains incomplete or sensitive to assumptions, AIMS must convert the residual uncertainty into concrete proposed evidence and directly test any invoked hidden variable, so that each pass closes by specifying the follow-up measurement or calculation that would most sharpen the conclusion.

\noindent\textbf{Navigation prompt and algorithms.} The navigation prompt specified all required MCP tools and parameters: 500 nm tip lift, $53\,\mathrm{\mu m}\times 53\,\mathrm{\mu m}$ scan size with $128\,$pixels$\times$$40\,$pixels, a $1636\,$pixels$\times$$1088\,$pixels optical image as the template, and termination criterion as 150 $\mu$m from chip center. To prevent the tip from straying outside the marker region after an inaccurate prediction, the maximum displacement between consecutive scans was limited to 100 $\mu$m. For Scan 3, GP reconstruction improved resolution from $128\,$pixels$\times$$40\,$pixels to $128\,$pixels$\times$$160\,$pixels.

Three algorithms were compared on the same set of scan images, with an "accurate" prediction defined as one lying in the same quadrant as the ground truth — sufficient to direct the subsequent move toward the sample. The convolutional neural network was fastest at inference but least accurate: it was trained on 5000 random $256\times256$ optical patches, each labeled by the arrow angle implied by the patch center relative to the map center ($\theta=\text{atan2}(y,x)$), for 30 epochs. It generalized poorly to actual scan images owing to differing noise levels, and training directly on scan data is impractical given the acquisition time per image. The NCC particle filter and the modern particle filter both treat localization as template matching on the optical image, with each particle proposing a map position whose cropped, rescaled patch is scored against the scan. The NCC method used 1000 particles over 10 iterations with single-metric scoring and minimal preprocessing beyond grayscale conversion, taking the final prediction from the dominant particle cluster (DBSCAN mode). The modern particle filter used a more robust pipeline: the reference image was preprocessed by row detrending, high-pass filtering, and gradient-based structural features; particles were initialized near promising locations from a coarse template-correlation step, then refined with a larger set (7500 particles, 25 iterations), each scored by a weighted combination of zero-mean NCC on high-pass patches, an unsigned gradient-orientation (HOG-like) term, and optional gradient-magnitude NCC, with a structure mask down-weighting flat regions. Because total navigation time is dominated by scanning rather than inference, we selected the modern particle filter to minimize wrong predictions and thereby save time overall.

\noindent\textbf{Evaluation of twist angle and GWC score.}
Before analysis, AIMS applied a third-order polynomial Savitzky–Golay filter with window size = 15 to smooth the target MIM spectrum to suppress high frequency noise. The most robust $\nu=1$ and $\nu=2$ features were first identified by assuming a sample twist angle range from $2^\circ$ to $4^\circ$ and searching the corresponding bottom gate voltage range. The local twist angle, filling factor and bottom gate voltage are connected via
\begin{equation}
    \theta\approx\sqrt{\frac{\sqrt{3}a^2}{2\nu}\frac{\varepsilon_0\varepsilon_r}{ed}(V_{bg}-V_\text{CNP})}
\end{equation}
where $a$ is the lattice constant of MoSe$_2$, $\nu$ is the filling factor, $\varepsilon_0$ is the vacuum permittivity, $\varepsilon_r$ is the hBN permittivity, $e$ is the unit charge, $V_{bg}$ is the gating voltage at $\nu$ and $V_\text{CNP}$ is the gating voltage at the charge neutrality point (CNP). The expression holds for all single-gated hexagonal twisted heterostructures (including graphene and other TMDs) in the small-angle limit, and can be extended to additional gating layers. The hBN thickness was supplied from AFM measurement; the lattice constant, unit charge, vacuum permittivity and hBN permittivity were retrieved by AIMS and were consistent with accepted values. $V_\text{CNP}$ was obtained by interpolating the linear $V_{bg}-\nu$ relation to $\nu=0$, leaving $\theta$ identifiable from $V_{bg}$ at a single assigned filling factor. 

Candidate dips were then identified as minima of the smoothed trace and, independently, as inflection features in its first derivative, the latter being essential for incompressible states that manifest as steps rather than isolated minima. A feature was retained only if its depth relative to a baseline defined by its flanking shoulders exceeded a significance threshold of twice the local noise, ensuring that only statistically meaningful features entered the analysis. The global CNP was determined by the average of $2V(\nu=1)-V(\nu=2)$ from an ensemble of curves with highest confidence in the $\nu=1$ and 2 dip extraction. The twist angle map at each grid point is then evaluated by this global CNP and local $V(\nu=1)$. 

To score the GWC states of each spectrum, AIMS focuses on the filling range $\nu \in [0.10,\ 0.90]$, and the score is assigned based on
\begin{widetext}
\begin{equation}
    S_i = N_{d} + (1-\varepsilon)\,\operatorname{minmax}\!\big(w_1\,\widehat{N}_{d} + w_2\,\widehat{\text{depth}}-w_3\,\widehat{\text{noise}} - w_4\,\widehat{\text{fillDev}}\big)
\end{equation}
\end{widetext}
where $\epsilon=0.001$ is the dominance margin, $\widehat{N_d}$ is the number of GWC dips, $\widehat{\text{depth}}$ is the added depth of all GWC dips, $\widehat{\text{noise}}$ is the noise level of MIM spectrum, $\widehat{\text{fillDev}}$ is the deviation from the expected fractional fillings $\nu=1/3, 1/2$ and $2/3$, and $w_{1,2,3,4}$ are the corresponding weights of each term with $w_{1,3,4}=1$ and $w_2=0.3$. An independent $N_d$ term is added to ensure a position with more GWC states observed always receives a higher score. 

\noindent\textbf{Melting temperature extraction.} For each spectrum, Savitzky–Golay filtering locates the $\nu=1/3, 1/2$ and $2/3$ states, and the temperature dependence of each dip depth serves as the melting diagnostic. Depth is referenced to a linear baseline fitted only to the dip-free shoulders (outer 28\% of each window edge), with the dip region excluded; when a state melts, the spectrum smooths and the shoulder-anchored line coincides with the central trace, so depth falls to $\sim0$, while the shoulder constraint prevents the baseline from bending into the dip and reporting spurious depth. Each depth is normalized by the difference between the 98th and 2nd percentiles of the smoothed spectrum to remove gain and offset variation. $T_m$ is defined as the temperature at which depth falls to 10\% of its maximum by linear interpolation. For $\nu=1/2$, which retains $\sim$48\% of its maximum depth at 20 K, $T_m$ is obtained by extrapolation with a shared power law $O(T)=A(1-T/T_c)^\beta$, with $\beta\approx0.7$ fixed by the in-range states.

\noindent\textbf{Sample design and fabrication.}
To encode directional information, we design lateral triangular Au markers whose acute angle always points toward the sample position. The markers cover a circular region 2 mm in diameter centered on the sample, spanning 48 distinct directions in total. Marker density increases toward the sample, providing a qualitative indication of relative distance to the target. Marker size was chosen so that multiple markers fall within a single scan frame. We additionally elongated the shorter side of the marker, which reduces the risk of recognizing the marker in the opposite orientation — a failure mode frequently encountered with our earlier narrower designs. We use standard electron beam lithography on poly(methyl methacrylate) (PMMA) to pattern the markers on the 285nm SiO$_2$/Si chip used for the sample locating test and deposit with Ti/Au (5/25 nm) by electron beam evaporation.

We exfoliate hexagonal boron nitride (hBN), graphite, and molybdenum diselenide (MoSe$_2$) flakes on O$_2$ plasma cleaned SiO$_2$/Si chip, and characterize the used flakes with optical microscopy and AFM. We adopt the dry transfer method for assembling the twisted MoSe$_2$ heterostructure. We prepare the standard transfer stamp with 6.5\% polycarbonate (PC) film on dome-shaped polydimethylsiloxane (PDMS). The first transfer is completed with a 15nm hBN pickup followed by graphite pickup, and final release onto a SiO$_2$/Si chip with pre-patterned Ti/Au (5/45 nm) markers and larger contacts. The PC film is washed off with remover PG, acetone and 2-propanol. The sample electrodes are then patterned and deposited with Ti/Pt (1/4 nm). The second transfer is completed in a glovebox with the monolayer MoSe$_2$ flake separated into two with an AFM tip and picked up sequentially after picking up a 6nm hBN. The stack is then melted down onto the previous hBN-graphite stack with sample electrodes. A final metal deposition is done with Ti/Au (5/45 nm) to make the electrodes for graphite and connections to outer electrodes. The twisted MoSe$_2$ region is cleaned with contact mode AFM to remove the bubbles. 

\noindent\textbf{MIM measurements.}
Microwave impedance microscopy (MIM) measurements are performed in a Janis $^3$He cryostat equipped with a 12 T superconducting magnet and customized scanning setup. The MIM probe is composed of an etched tungsten tip with 50 nm apex mounted on a quartz tuning fork, which enabled topographic feedback. The tip is maintained around 10-30 nm away from the sample surface during the measurement. 

The control of gating voltage, scanning, and tip movement is achieved with Nanonis SPM controller (SPECS Group, RC5). The scanners and positioners are assembled and controlled with commercial piezoelectric actuators and piezo positioning controller (Attocube, ANC300). 

The microwave frequency is set to 1GHz with input power about 0.2 $\mu$W using a signal generator (Agilent, E8257D), and no magnetic field is applied for the discussed measurements. To reduce the Schottky barrier between Pt contact and twisted MoSe$_2$, we apply 750 nm light using a supercontinuum laser (NKT Photonics, FIU-20), with wavelength selected by a monochromator (Princeton Instruments, Acton SP 2300).

\noindent\textbf{Open-Source Software}
AIMS acknowledges the following open-source resources: the Model Context Protocol Python SDK (FastMCP), LibreChat, Starlette, Uvicorn, Pydantic, NumPy, SciPy, pandas, Matplotlib, scikit-learn, scikit-image, OpenCV, Pillow, pySPM, PyVISA (with the pyvisa-py backend), pylablib, NanonisTCP, and PyYAML.

\section*{Acknowledgements}

We thank the help from K. Lei in reviewing the manuscript.

We acknowledge the support of Panofsky fellowship at SLAC national accelerator laboratory, and start up fund from Boston College. The works at the Stanford Institute for Materials and Energy Sciences are supported by the U.S. Department of Energy, Office of Science, Office of Basic Energy Sciences, Division of Materials Sciences and Engineering, under Contract No. DE-AC02-76SF00515. The works at University of Florida are supported by the U.S. Department of Energy, Office of Science, Office of Basic Energy Sciences, Under Contract No. DE-SC0022216. 

\section*{Competing interests}

The authors declare no competing interests.

\section*{Additional Information}

Correspondence and requests for materials should be addressed to the corresponding author.

\section*{Data and Code Availability}
All data and code used in this study are available from the corresponding authors upon request.

\bibliographystyle{naturemag}
\bibliography{refs}

\begin{figure*}
    \centering
    \includegraphics[width=1\linewidth]{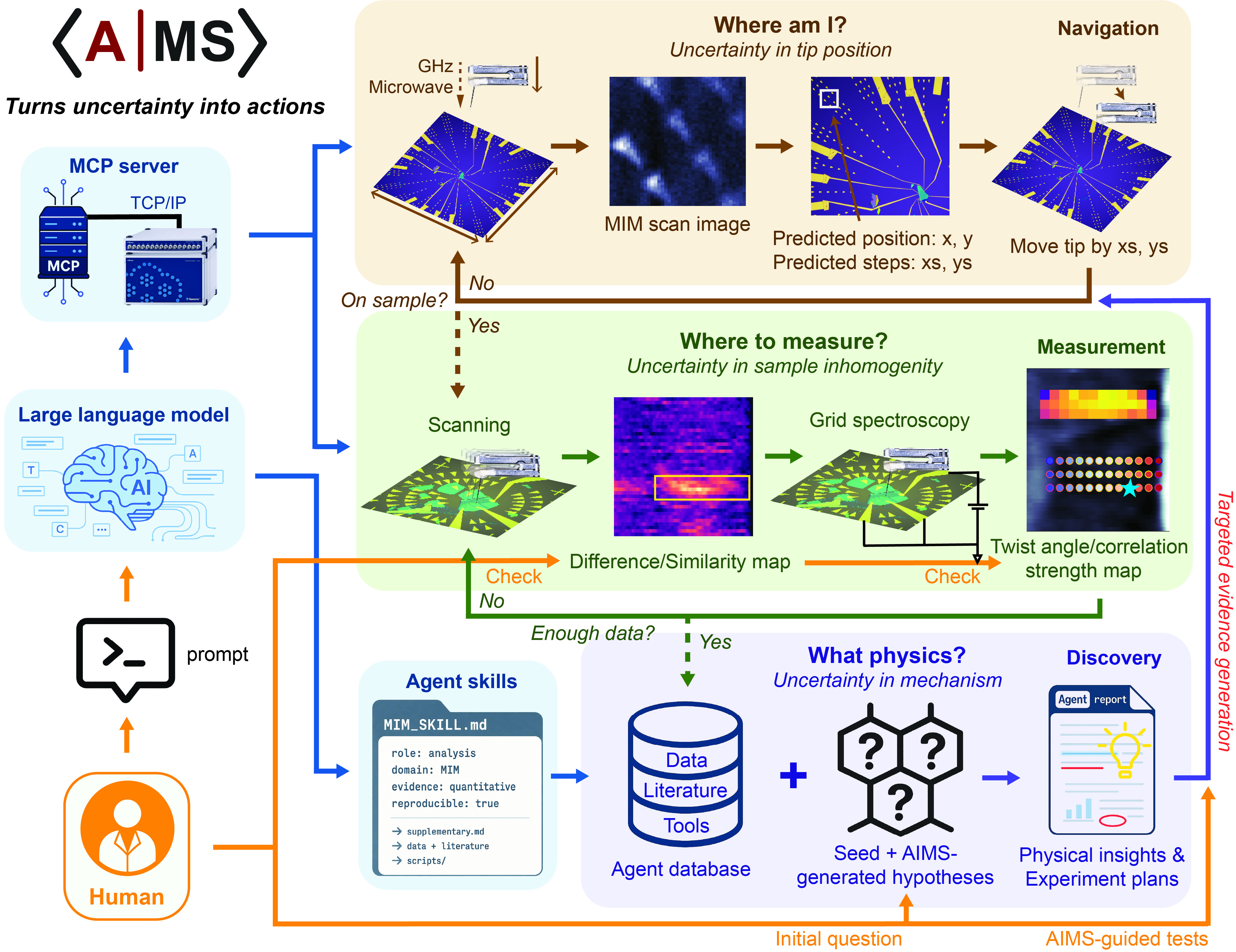}
    \caption{\textbf{AIMS turns three generic experimental uncertainties into action through three nested loops.} AIMS converts uncertainty at three stages of a cryogenic MIM experiment into targeted actions. A human researcher initiates a task prompt through large language model interface, after which AIMS calls the requisite instrument control and real-time data analysis tools from the MCP server as each loop proceeds. In the navigation loop, it acquires MIM images, estimates the tip position and confidence, moves toward the sample, and invokes recovery procedures when localization is unreliable. In the measurement loop, large-area gate-dependent imaging and grid spectroscopy map sample inhomogeneity and rank candidate sites according to twist angle, disorder, and correlated-feature quality, guiding the experiment toward the most informative position. In the discovery loop, researchers provide the scientific question and initial seed hypotheses, while AIMS generates additional physical, hidden-variable, and measurement-artifact hypotheses and ranks the surviving candidates against the available evidence. When the evidence is insufficient, AIMS identifies the limiting gap, performs targeted theoretical calculations, or guides registered follow-up measurements. The new evidence is returned to the same inference framework, allowing AIMS to update both the mechanism ranking and the role of each hypothesis. The arrows and components in blue denote the internal agent stream. The orange arrows denote researcher input, physical execution, and optional oversight.
}

    \label{fig:1}
\end{figure*}

\begin{figure*}
    \centering
    \includegraphics[width=1\linewidth]{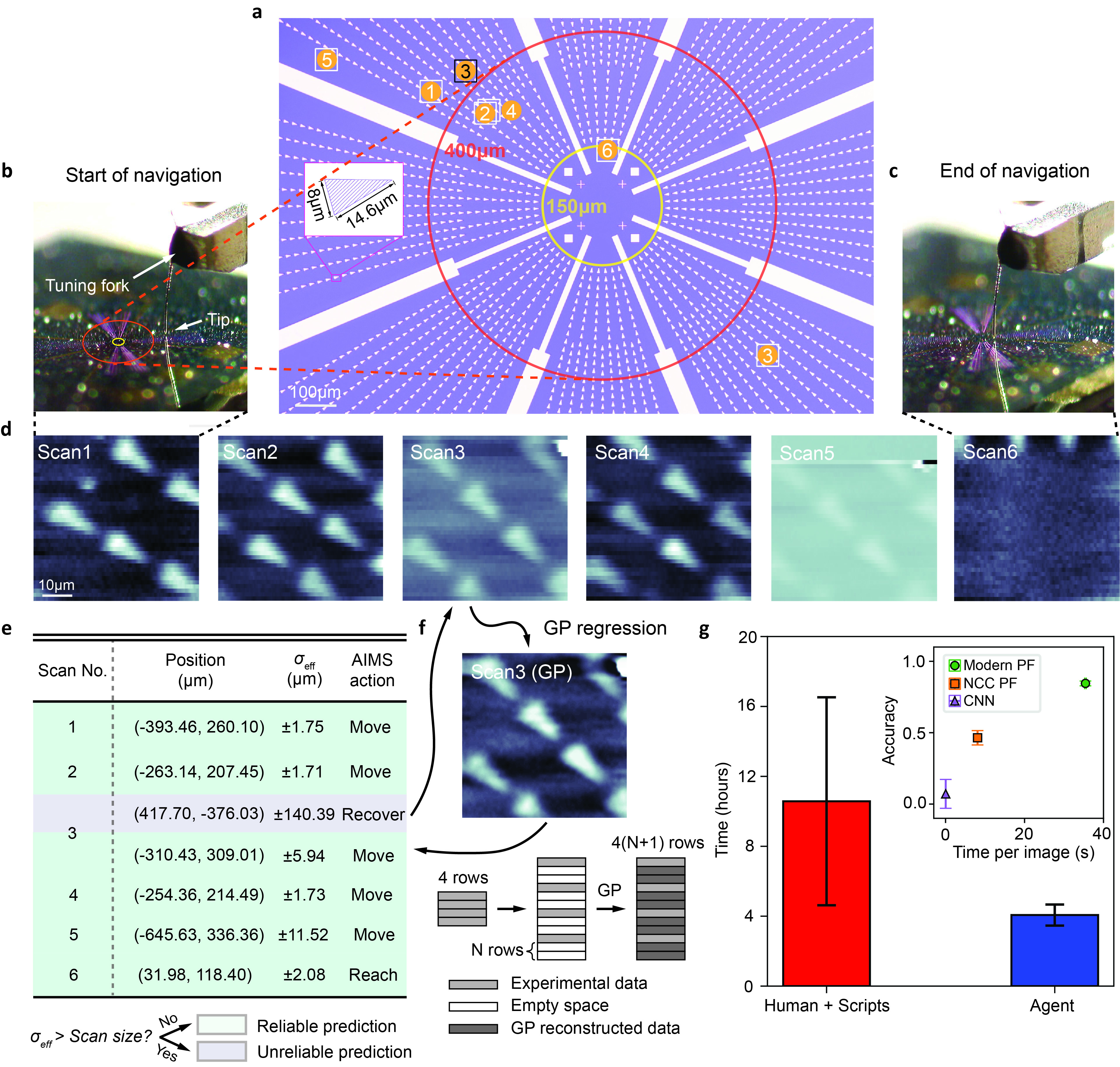}
    \caption{\textbf{AIMS locates the sample under cryogenic drift and recovers from unreliable position estimates.} (a) Optical image of the marker chip. The white squares in numerical order mark the predicted positions of MIM scans taken at each step. The black square marks the predicted position of Scan 3 after GP regression reconstruction. The yellow and orange circles highlight 150 $\mu$m and 400 $\mu$m from the pattern’s center respectively. The inset shows the design geometry of the lateral triangular arrow marker. (b) \& (c) Optical image of the tip with respect to the marker pattern at initial position after approaching (Scan 1) and final position after the sample locating process ends (Scan 6) respectively. (d) MIM-Im signal for each scan. The scan frame for each image is $53~\mu$m $\times$ 53 $\mu$m. The change of background color in scan 3 and 5 are caused by the retraction of tip after touching defects on the chip, which dramatically reduce the MIM signal. (e) Predicted positions and effective uncertainty $\sigma_\text{eff}=\sqrt{\sigma_x^2+\sigma_y^2}$ for each scan image in (d) based on the template image. The reliability of prediction is determined by comparing $\sigma_\text{eff}$ to the $53~\mu$m scan size. The initial highly uncertain prediction for Scan 3 triggers recovery process of AIMS, leading to enhanced prediction with GP regression reconstruction of the same scanning image. (f) Top image shows the GP regression reconstructed Scan 3 image with 4 times better resolution. Bottom image shows a demonstration of GP regression reconstruction on a scanning image with 4 rows of data. The 4 rows of data are separated by $N$ empty rows and proceeded with GP regression to fill the empty rows with reconstructed data, leading to $4(N+1)$ rows of data and hence enhanced resolution. (g) Total time spent in sample navigation for human equipped with scripted codes-driven and AIMS-driven experiment, including scanning, tip movements, and position prediction. The top right inset shows the accuracy and time spent for modern particle filter (PF) method, normalized cross correlation (NCC) particle filter method, and convolutional neural network (CNN) method in predicting the position of one scanning image.
    }
    \label{fig:2}
\end{figure*}

\begin{figure*}
    \centering
    \includegraphics[width=1\linewidth]{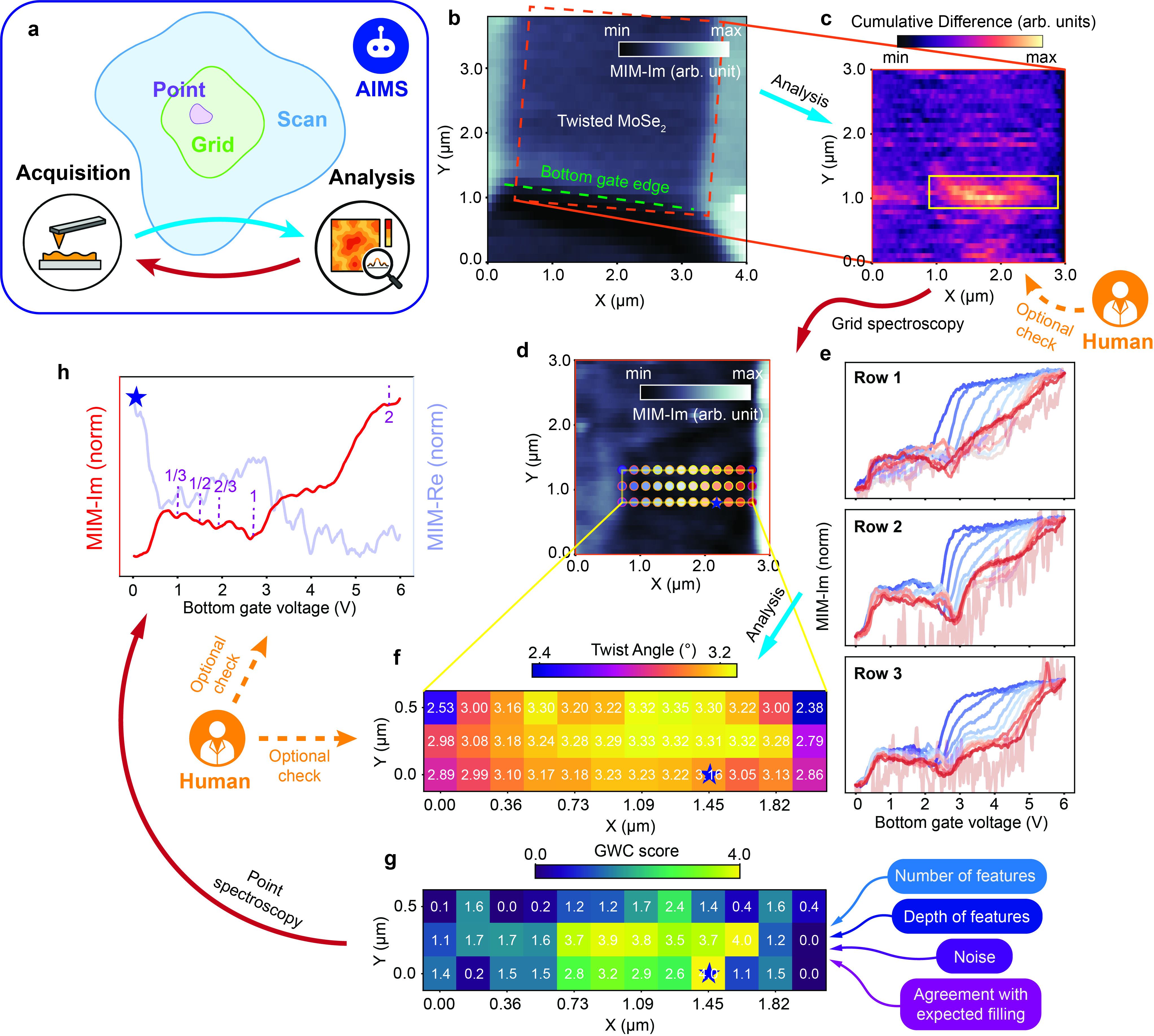}
    \caption{\textbf{AIMS selects measurement sites in a disordered twisted MoSe$_2$ device and resolves fractional generalized Wigner crystal states.} (a) Schematic of AIMS-driven measurement loop. The process iterates between data acquisition and analysis, progressively narrowing the region of interest from a large scan area to a smaller grid and ultimately to the single optimal measurement point, thereby improving experimental efficiency. Red and blue arrows in the remaining panels denote the acquisition-to-analysis and analysis-to-acquisition directions, respectively. The orange dashed lines coupled to human represents the optional checks that human researcher can perform. (b) MIM-Im scan on the twisted MoSe$_2$ sample region. The green dashed line marks the edge of the bottom graphite gate. The orange rectangle marks the scan region in (c) taken by AIMS. (c) Cumulative difference map on a smaller region in twisted MoSe$_2$ calculated from MIM-Im scan taken by AIMS at bottom gate voltage from 2.5 to 3.5 V. The yellow rectangle marks the agent determined best region for grid measurement. (d) MIM-Im scan on the same region in (c). The dots represent the measurement positions of point spectroscopy in (e) with corresponding color, and the color of dot border represents the twist angle extracted in (f). The blue star marks the agent determined position with strongest fractional correlated features. (e) MIM grid spectroscopy taken by AIMS at the dotted positions in (d) with respect to bottom gate voltages. The color from blue to red represents position from left to right. Row 1 matches the top most row in (d). The measurement at right most column is largely screened by the contact, and is thus more noisy and plotted with reduced opacity. (f) Twist angle distribution map of the grid in (d) derived by AIMS. The local twist angle is extracted with the MCP server based on the charge neutrality point and the most prominent $\nu=1$ feature. (g) GWC score map of the grid in (d) evaluated by AIMS. The score is determined based on number of fractional correlated features, depth of features, noise, and consistency with expected fractional filling. (h) MIM point spectroscopy taken by AIMS at the blue star position. The spectrum is smoothed by a Savitzky–Golay filter with window size = 15 and polydeg = 3. The filling factors in purple are assigned with both human and agent inputs.
    }
    \label{fig:3}
\end{figure*}

\begin{figure*}
    \centering
    \includegraphics[width=1\linewidth]{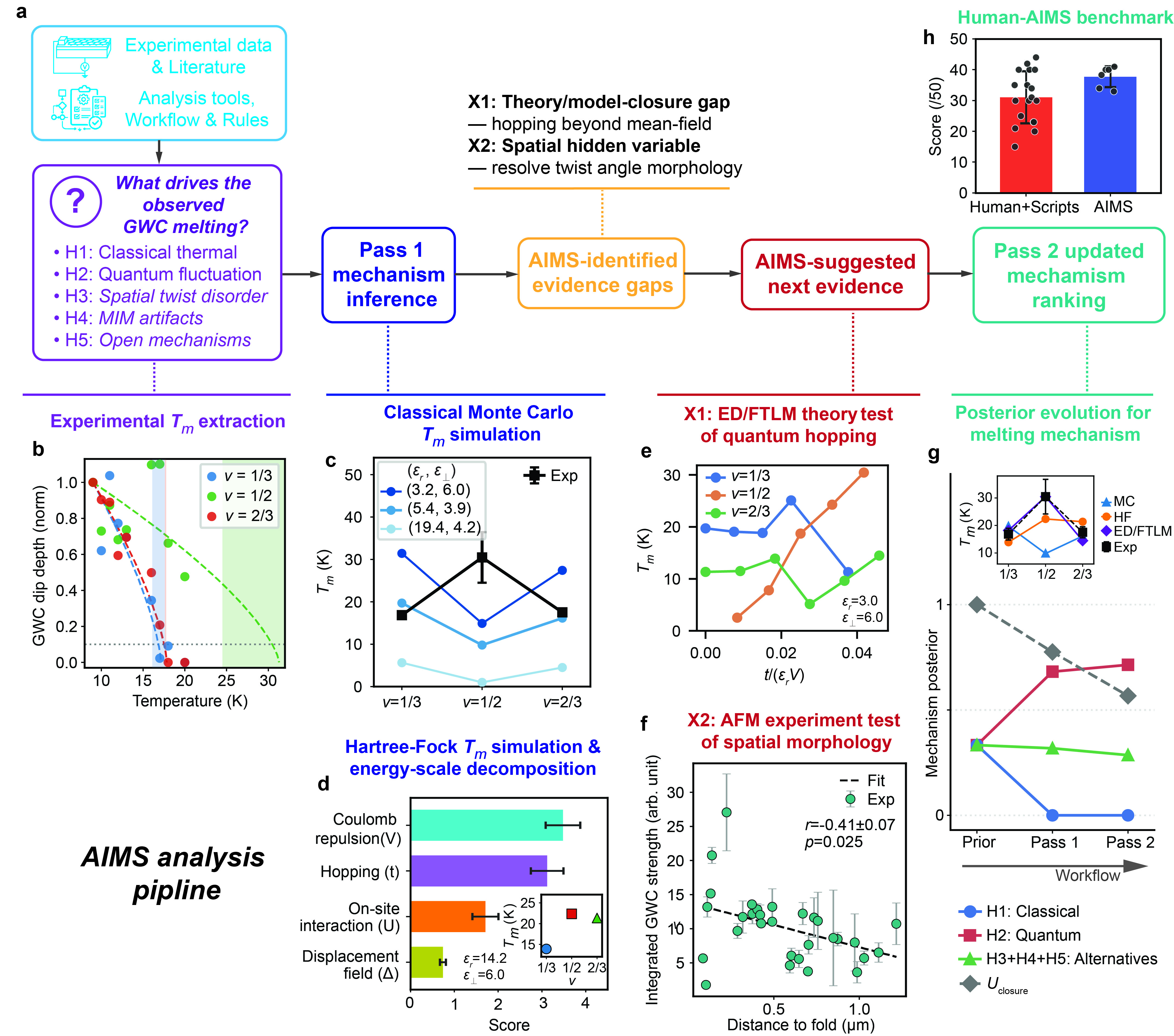}
    \caption{\textbf{A closed AIMS discovery loop traces the anomalous stability of the half-filled generalized Wigner crystal to quantum fluctuations.} (a) Discovery workflow of AIMS. A curated database of literature \cite{barber_microwave_2021,Huang2021FractionalWS2WSe2,Huang2025CorrelatedInsulatorGaps,Ji2023HarnessingExcitons,Li2021ImagingWignerCrystals,Yang2025CorrelatedInsulating,Wang2023ChernInsulatorMIM,Spivak2004IntermediatePhasesWigner,Kumar2026MeltingTemperatureGWC,zong2025quantum, Kumar2025OriginWignerCrystallinity} and experimental data, together with data analysis tools, workflows and rules as MIM discovery skills, is supplied to AIMS prior to each run. A scientific question is posed by human to AIMS as prompt alongside the physics priors and hypotheses (H1 -- H2). With these information, AIMS adds its new hypotheses (H3 -- H5), performs the first pass of analysis and score each hypothesis with Bayesian inference. To improve the credibility of its judgment, AIMS identifies the unresolved evidence gaps and proposes follow-up theoretical and experimental tests. The suggested test results are reanalyzed by AIMS as the second pass, which updates the ranking of hypotheses with enhanced confidence. (b) The normalized GWC dip depth for filling factors $\nu=1/3, 1/2$ and $2/3$ from 9 to 20~K. The black dashed line denotes the 10\% threshold. The colored dashed lines are fits to the corresponding GWC dip depth as a function of temperature. The shaded areas indicate the uncertainties in the extracted $T_m$ for each filling. (c) $T_m$  obtained from classical Monte Carlo simulations for several choices of the relative permittivity $\varepsilon_r$ and $\varepsilon_\perp$, compared with experimental $T_m$ from (b). (d) Importance of each energy scale in determining $T_m$ across all GWC states, simulated based on the Hartree-Fock model with Bayesian optimized parameters $\varepsilon_r=14.2$ and $\varepsilon_\perp=6.0$. The inset shows the corresponding $T_m$ for each filling. (e) $T_m$ of each GWC state as a function of $t/(\varepsilon_rV)$, where $t$ is the hopping amplitude and $V$ the Coulomb repulsion, obtained from exact diagonalization and finite-temperature Lanczos method at $\varepsilon_r=3$ and $\varepsilon_\perp=6.0$. (f) The correlation between integrated GWC strength of the grid measurement and the distance to observed fold. The Pearson coefficient is fitted to be $-0.41\pm0.07$ with $p$-value $0.025$. (g) Evolution of melting mechanism posterior with respect to the addition of new evidence in each pass. The quantum-fluctuation driven mechanism overwhelms other mechanisms with all evidence from pass 1 and 2. Alternatives represents a weighted average of mechanism hypotheses H3 to H5. $U_\text{closure}$ represents the model-closure uncertainty in the AIMS framework. The inset shows the optimized $T_m$ from MC, HF, ED/FTLM models compared to the experimental $T_m$ for each filling.  (h) Blind scores (out of 50) assigned by three condensed-matter physicists to GWC melting analysis reports composed by the agent versus by humans from the same experimental dataset.
    }
    \label{fig:4}
\end{figure*}

\end{document}